\documentclass[amsmath,prl,aps,twocolumn]{revtex4}%
\usepackage{amsthm,amsfonts,graphicx,verbatim}
\usepackage{hyperref}

\newcommand{\be}{\begin{equation}}
\newcommand{\ee}{\end{equation}}
\newcommand{\bea}{\begin{eqnarray}}
\newcommand{\eea}{\end{eqnarray}}

\renewcommand{\vec}[1]{{\bf #1}}
\renewcommand{\epsilon}{\varepsilon}
\def\nn{\nonumber\\}

\usepackage{color}
\definecolor{Red}{rgb}{0.8,0,0}
\definecolor{Black}{rgb}{0,0,0}

\begin{document}
\title{Full counting statistics and the Edgeworth series for matrix product states}
\author{Yifei Shi and Israel Klich}
\affiliation{Department of Physics, University of Virginia, Charlottesville, VA 22904}
\begin{abstract}
We consider full counting statistics of spin in matrix product states. In particular, we study the approach to gaussian distribution for magnetization.  We derive the asymptotic corrections to the central limit theorem for magnetization distribution for finite but large blocks in analogy to the Edgeworth series. We also show how central limit theorem like behavior is modified for certain states with topological characteristics such as the AKLT state.\end{abstract}
\maketitle

%\section{Introduction} 
With the advantage of precision experiments in condensed matter physics, it is now possible to probe the nature of correlated quantum systems to ever higher degrees of detail and precision. This is especially apparent in cold atom systems, where in addition to refined correlation measurements, we also have fine control of the Hamiltonian parameters themselves. One of the most useful ways of studying detailed correlations in the intermediate regime between the macroscopic, thermodynamic properties, and the microscopic, atom by atom level, is through the full counting statistics functions. These describe the full probability distribution of suitable observables, such as the magnetization of a block of spins in a spin chain or the total excess charge flowing through a quantum point contact.

The full counting statistics function (FCS) contains detailed information about the properties of the state. 
It has been a useful tool to analyze quantum states, from it's original appearance in quantum optics, in the theory of photon detectors \cite{glauber1963photon,mandel1965coherence} in quantum 
optics to counting statistics of electrons in mesoscopic systems introduced by Levitov and Lesovik \cite{levitov1992charge}. The full counting statistics has studied in numerous electronic systems theoretically  \cite{nazarov2003full,belzig2005full,schonhammer2007full,komnik2011quantum,Levine2012Bantegui} as well as in experiments \cite{bomze2005measurement,flindt2009universal}. The utility of full counting statistics for cold atoms was pointed out in \cite{cherng2007quantum}. It has also been demonstrated that in certain cases counting statistics may be used to characterize block entanglement entropy in fermions states and spin states \cite{klich2006measuring,klich2009quantum,song2011entanglement}. 
Recently, the analyticity properties of the "bulk" component of the full counting statistics in classical Ising and quantum XY spin chains has been used as an alternative characterization of phases \cite{ivanov2012characterizing}.% (\new more?  \old) 

Here, we explore the quantum noise in an important class of 1d states, the so called matrix product states (MPS) \cite{Finitelycorrelatedstates}. Such states appear as the ground states of frustration free Hamiltonians, and are especially useful as variational states to study other 1d quantum systems.
%Here we concentrate on the behavior of full counting statistics function of block magnetization and it's scaling behavior for large blocks of spins. 
%Variational wave functions are one of the most powerful tools in the arsenal of a physicist, useful for establishing various bounds on energetics and for describing other features of systems of physical interest. In recent years, a number of proposed variational states have been popular due to their natural simple form, which hopefully captures ground state properties of realistic Hamiltonians. Among these are matrix product states (MPS)\cite{Finitelycorrelatedstates}, projected entangled pairs (PEPS)\cite{Schuch07} and other tensor network states, MERA, RAGE and more \cite{RAGE2011}. 
MPS posses convenient properties that allow a thorough study of correlations and fluctuations in them,  for example, an analogue of the Wick's theorem has been demonstrated for generic translationally invariant MPS in \cite{hubener2012wick}. Finally, we remark that our results also hold for non quantum states as long as the probability distribution of certain measurements may be described in terms analogous to MPS, a prominent example for which is the exact solution of the 1D asymmetric exclusion process \cite{derrida1999exact}, in a recent paper, the full counting statistics for the asymmetric exclusion model was considered in \cite{gorissen2012exact}.

Here, we concentrate on the corrections to central limit theorem (CLT) like behavior of the full counting statistics. The central limit theorem is a description of the statistics of averages of independent random variables, stating that properly weighted average tend to a Gaussian distribution when the number of random variables is large. Since the correlation length in a MPS, say, is finite, one may expect gaussian like behavior for the magnetization of large blocks of spins. We find that the simple structure of the MPS allows us to not only do this but much more: we can controllably identify how the central limit of magnetization is reached, what are the main corrections (a consequence of entanglement in the system) and show how CLT may sometimes completely fail in cases of topological states.

The distribution of magnetization approach to Gaussian behavior at large spin blocks is substantially more intricate for the MPS as opposed to independent random variables. To address this behavior we concentrate on deriving the asymptotic probability distribution and corrections to it. While in many cases, even when the distribution seems Gaussian in the infinite block size limit, the corrections due to finite block size are modified. Such corrections, are described, for independent, identically distributed  variables using various asymptotic series such as the Gram-Charlier A series and the Edgeworth series \cite{EdgeworthKolassa,Edgeworth}. The Edgeworth series has been extensively studied in the mathematical literature, with the focus on ensuring it's applicability when dealing with random identically distributed variables, which may have divergent moments, see e. g.  \cite{esseen1945fourier,petrov1964local,bobkov2011rate}.

Here, we derive an expression for the asymptotic corrections to the central limit error function of MPS in analogy to the asymptotic Edgeworth series. We start by deriving formulas for the full counting statistics generating function. 
%\section{Full counting statistics for Matrix Product States}
For an extensive review of MPS their properties, and their relation to DMRG see \cite{schollwock2005density,McCulloch07MPS}. MPS have been shown to be  extremely well suitable for approximation of ground states of gapped Hamiltonian \cite{Hastings07,Verstraete06}.
Let us consider a MPS with periodic boundary conditions on $N$ spins, defined as follows: 
\bea \left.\psi =\Sigma _{\{\sigma \}}\text{Tr} \left(A_{\sigma _1}A_{\sigma _2}\text{..}A_{\sigma _N}\right)|\sigma _1\sigma _2\text{...}\sigma _N\right\rangle
\eea
 {where }
$
A_{\sigma }\in \left\{A_1,\text{..}A_S\right\}
$
, $S$ is the spin index, and $\sigma _1\in \{1,\text{..}S\}$. The matrices A are of size $D\times D$,  where $D$ is called the bond dimension.

To express the full counting statistics of the spin variable $\sigma$, we define:
\bea\label{def of E(lambda)}
E(\lambda)=\Sigma _{\sigma }e^{i \lambda  \sigma  }\bar{A}_{\sigma }\otimes A_{\sigma }.
\eea
The full counting statistics generating function of the magnetization of a block of $l$ sites is then given by
\bea\label{counting statistics chi}&
\chi (\lambda ;l;N)\equiv \sum_n \text{prob}(\text{total~spin~of~block}=n) e^{i\lambda n}=\nn &
=\left\langle e^{i \lambda \hat{\cal S}_l }\right\rangle =\frac{\text{Tr} E(\lambda )^l E(0)^{N-l}}{\text{Tr} E(0)^N}.
\eea
Here $\hat{{\cal S}}_l =\Sigma _{i=1}^l \hat{\sigma}_i$ where $\hat{\sigma}_i$ is a spin operator at site $i$. %, with eigenvalues $\in\{1,S\}$.
When considering the thermodynamic limit, we add, as is usual, the demand that:
\bea
\Sigma_{\sigma=1}^SA_{\sigma }A_{\sigma }^+=I
\eea
This ensures that the largest eigenvalue of $E(0)$ is $\lambda=1$. In addition, when dealing with problems in the thermodynamic limit we assume this largest eigenvalue is non degenerate. %, with left and right eigenvectors $\langle L(0)|$ and $| R(0) \rangle$.
In the thermodynamic limit, we define:
\bea
\chi (\lambda ;l)\equiv \lim _{N\to \infty } \chi (\lambda ;l;N)%\left\langle e^{i \lambda  \Sigma _{i=1}^l \sigma _i}\right\rangle 
\eea
The existence of the limit is assured by the conditions above. To compute $\chi (\lambda ;l)$ we define $P(\lambda)$ to be the matrix which brings $E(\lambda)$ to it's Jordan form, with Jordan blocks $J_k$ arranged such that the block with the largest eigenvalue is $J_1$. Note that the number of blocks as well as eigenvalues depend on $\lambda$. We have:
\bea
E(\lambda)^l=P(\lambda)(\oplus_{k=1}J_k^l)  P^{-1}(\lambda)
\eea
We note that if there is no degeneracy,  
\bea
E(0)^N\rightarrow P(0) |1\rangle\langle 1|P^{-1}(0)
\eea
Therefore the full counting statistics function is given by:
\bea \chi (\lambda ;l)=\langle 1| P^{-1}(0)P(\lambda)(\oplus_{k=1}J_k^l)  P^{-1}(\lambda) P(0) |1\rangle   \eea
Let $\alpha_k(\lambda)$ be the diagonal value of $J_k$. We can compute explicitly the power of a Jordan block, obtaining after some algerba the formula:
\bea\label{FCS general form}
\chi (\lambda ,l)=\sum_{k=1}\sum_{i=1}^{d_k}\sum_{\nu=0}^{min(l,d_k)-i} C_{\nu }^l Q_{k,i,\nu } \alpha_k^{l-\nu } 
\eea
where $C_{\nu }^l$ are the binomial coefficients, $d_{k}$ the dimension of Jordan block $k$ and
\bea &\label{Q coefs}
Q_{k,i,\nu }(\lambda)= \langle i+\nu+\sum_{n=0}^{k-1}d_n|P^{-1}(\lambda) P(0) |1\rangle \times \nn & \langle 1| P^{-1}(0)P(\lambda)|  i+\nu+\sum_{n=0}^{k-1}d_n\rangle  .
\eea
Note that in \eqref{FCS general form},  $Q, \alpha_k$ as well as the limits in the sum depend on $\lambda$ implicitly. 

Since the largest eigenvalue 1 of $E(0)$ is non-degeneracy, we have that $Q_{1,1,0}(0)=1$ and $Q_{k,i,\nu}(0)=0$ for all other value of $k,i,\nu$. 
It is also important note that since, generically, eigenvalues do not cross, we expect that we may set $d_{k}=1,\nu=0$ in \eqref{FCS general form} for almost all values of $\lambda\in [-\pi,\pi]$, unless some special symmetry or constraint is present.

%Next, we break up $\chi(\lambda)$ as:
%\bea
%\chi(\lambda,l)=\chi_1(\lambda,l)+\delta \chi(\lambda,l)
%\eea
%where in $\chi_1(\lambda)$ we put the contribution from the largest eigenvalue, which we for simplicty assume non degenerate for $\lambda\in[0,2\pi]$. It is interesting to while $\chi_1$ does not describe a probability distribution, it describes a distribution with related features.

%, giving us:
%\bea \chi (\lambda ;l)=\left\langle L(0)\right|E(\lambda )^l\left|R(0)\right\rangle.  \eea

Let us now consider the limit of large block size $l$. As with any thermodynamic quantity, computed in a system with finite correlation length, we expect a gaussian distribution of observables according to the CLT. %, with certain finite size corrections. %The CLT  certainly holds if no spin correlations are present. %in which case we will have 
%\[ log(\chi(\lambda,m)) = m~ log( \chi(\lambda,1)) .\]  
%If we then measure the probability distribution of properly normalized magnetization, it will converge to a Gaussian form when $m$ is large.
For a matrix product state, of course, the spins are not independent, and so the central limit distribution receives contributions from two types of corrections: due to correlations and due to finite size. % However, since the system has a finite correlation length and (approximate) size extensivity of the  block magnetization, we may therefore still expect the probability distribution to be Gaussian. 
Bellow we establish this behavior and derive the appropriate asymptotic description of the probability distribution for large but finite blocks. \\

In the limit of large $l$, if the largest eigenvalue of the matrix $E(\lambda )$ is non degenerate, 
$E(\lambda )^l$ is  dominated by the largest eigenvalue $\alpha _{1}(\lambda )$  and %the associated eigenvectors 
%$\langle L_{max}(\lambda)|$ and $|R_{max} (\lambda)\rangle$ of $E(\lambda)$. If we assume $E(\lambda)$ is diagonizable, we have:
%\bea
%\chi (\lambda ;l)\to \left \langle L(0)|R(\lambda )\rangle \langle L(\lambda)|R(0 )\right\rangle \alpha_{\max}(\lambda)^l
%\eea
we may write that: 
\bea\label{chi asymptotics}
\chi (\lambda ;l)\sim\chi _0(\lambda )\chi_1(\lambda)^l~~~~as~~~l\rightarrow\infty\eea
where:
$$
\chi _1(\lambda )=\alpha _{1}(\lambda )~~ ;~~\chi _0(\lambda )=Q_{1,1,0}(\lambda)$$
% \langle L(0)|R(\lambda )\rangle \langle L(\lambda)|R(0 )\rangle.
It is possible to take into account the corrections due to the smaller eigenvalues of $E(\lambda)$ as well, giving additional exponentially small corrections. %As we show later in the paper, these smaller eigenvalues may become of interest when dealing with interfaces.

We are now in position to describe the probability distribution of block magnetization. Let us define: 
\bea\label{def M_l}
\hat{M}_l=\frac{1}{\sqrt{l}}\frac{\hat{\cal S}_l-l \mu(l)}{var(\sigma,l)}
\eea
where $\mu(l)$ is the average magnetization per site, and $var(\sigma,l)$ is the variance {\it per site}.  We note that since the spin variables on different sites are not independent,  both  $\mu(l)$ and $var(\sigma,l)$ depend explicitly on the size of the block. Let \bea\label{Def F_l}
F_l(M)=Prob(M_l\leq M)
\eea
be the probability distribution of measuring $\hat{M}_l$. To find $F_l(M)$, we now focus on the % $l$ dependence of $\chi(\lambda;l)$, when $l\rightarrow \infty$, the
 FCS for $\hat{M_l}$, defined as:
\bea \chi_M(\lambda;l)= \langle e^{i\lambda \hat{M}_l } \rangle. \eea
Since we assume that the largest eigenvalue of $E(\lambda)$ is non degenerate at $\lambda=0$, this eigenvalue is analytic in a neighborhood of $\lambda=0$. Indeed both  $\chi _0(\lambda ),\chi _1(\lambda )$ are analytic in the domain $\lambda\leq |\lambda_*|$ where $\lambda_*$ is the smallest $\lambda$ (in the complex plain) for which the largest eigenvalue of $E(\lambda)$ becomes degenerate (see, e.g.  \cite{lax1996linear}).  
We can therefore expand $log(\chi_1(\lambda)),log(\chi_0(\lambda))$ near $\lambda=0$. Noting that $\chi _0(\lambda )=\chi _1(\lambda )=1$ we have the "cumulants"  $\kappa_r $   and $\xi_r $ in:
\bea
log(\chi_0(\lambda)) &=& \sum_{r=1}^{\infty}\frac{\xi^{r}(i\lambda)^r}{n!}  \nonumber\\
log(\chi_1(\lambda)) &=& \sum_{r=1}^{\infty} \frac{\kappa^{r}(i\lambda)^r}{n!}.
\eea
We see that for a block of $l$ spins, $\mu(l)=\langle\sigma\rangle = \kappa_1+\xi _1/l$, and $var(\sigma,l)=\sqrt{\langle \sigma^2 \rangle-\langle \sigma \rangle^2} = \sqrt{\kappa_2 + \xi_2/l}$. We also recognize $\chi_0$ as boundary (or ``Edge") term that characterize the effect of the rest of the chain on the chosen $l$ spins, and $\chi_1$ as the bulk term that is not effected by other  spins. \\

In this paper $\chi_1$ plays, formally, the role of the local independent random variable in the usual derivation of the central limit theorem. However, it is important to note that in a generic MPS, $\chi_1$ is not the full counting statistics of a valid probability distribution. Indeed, for that, the associated distribution, given by the Fourier transform of $\chi_1$ must be a positive real function. 

Let us briefly explore how close $\chi_1$ is to a valid probability distribution. To do so we write the counting statistics of a block of size 1 as:
\bea
\chi(\lambda,1)=\chi_1(\lambda)+ \chi_{\delta}(\lambda),
\eea 
and define the pseudo-probabilities:
\bea\label{pseudo probs}
\tilde{p}_{n,i}=\frac{1}{2 \pi }\int _{-\pi}^{\pi }\text{d$\lambda $} ~ \chi_i\left(\lambda \right) e^{-i \lambda  n} ~~;~~i=1,\delta.
\eea
Associated with the various eigenvalues of the $\chi_1(\lambda),\chi_{\delta}(\lambda)$. Thus $\tilde{p}_{n,1}$ is the effective probability distribution which would have generated the asymptotic behavior described by the central limit behavior. We can establish the following properties:

1) From the definition of $\chi_1$, we can immediately infer that the associated distribution is discrete. %and real (however it is in general not necessarily positive).

Indeed, observe that since $E(\lambda)$ is periodic, we can choose  $\chi_1(\lambda)$ to be periodic: $\chi_1(\lambda)=\chi_1(\lambda+2\pi)$, which is associated with discrete, integer, spins. 

2) The distribution is real (however it is in general not necessarily positive).

The second property is established by noting that: 
\bea
E(-\lambda)=\tau \overline{E(\lambda)}\tau
\eea
where $\tau$ swaps $v_1\otimes v_2\rightarrow v_2\otimes v_1$. Therefore 
\bea
spec E(-\lambda)=spec \overline{E(\lambda)}
\eea
and, in particular $\chi_1(-\lambda)=\chi_1^*(\lambda)$, ensuring the Fourier transform is real.
It is, however, in general not associated with a probability distribution, since the Fourier transform is, in general, not strictly positive.

3) $\sum_{n}\tilde{p}_{n,1}=1$ and $\sum_{n}\tilde{p}_{n,\delta}=0$. To prove this note that $\chi_\delta(0)=0$ since $\chi(0,1)=\chi_1(0)=1$. Now use that $\sum_{n}\tilde{p}_{n,\delta}=\chi_\delta (0)$.

We now proceed to derive the Edgeworth series for our MPS.
Using the definition \eqref{def M_l} and eq. \eqref{chi asymptotics}, we can find the cumulants for the distribution of $\hat{M_l}$:
\begin{align}&
log\chi_M(\lambda;l) = \frac{(i\lambda)^2}{2} + \frac{(i\lambda)^3 (l\kappa_3+\xi_3)}{6 (l\kappa_2+\xi_2)^{3/2}} \nonumber
\\ &+\frac{(i\lambda)^4 (l\kappa_4+\xi_4)}{24 (l\kappa_2+\xi_2)^2}+\frac{(i\lambda)^5 (l\kappa_5+\xi_5)}{120 (l\kappa_2+\xi_2)^{5/2}} + ... 
\end{align} 

We note that for the normal distribution,  we have:
\bea
log(\phi(\lambda)) = \frac{(i\lambda)^2}{2} \eea
Combine the two equations, and collect terms according to the power of $l$, we have 
\begin{align}&
log\frac{\chi_M(\lambda)}{\phi(\lambda)} = \frac{1}{l^{1/2}}\frac{(i\lambda)^3\kappa_3}{6\kappa_2^{3/2}}+ \frac{1}{l}\frac{(i\lambda)^4\kappa_4}{24\kappa^2} \nonumber
\\&+ \frac{1}{l^{3/2}}[\frac{(i\lambda)^5\kappa_5}{120\kappa_2^{5/2}}+\frac{(i\lambda)^3}{6}(\xi_2-\frac{3\xi_3}{2\kappa_2})]+...
\end{align}
Exponentiate the above equation, we have
\bea
\chi_M(\lambda;l) = (1+\sum_{j=1}^\infty \frac{q_j(i\lambda)}{l^{j/2}} )e^{-\lambda^2/2}
\eea
where $q_j$ is a polynomial of degree $3j$. 

Finally, to obtain $F_l$ in \eqref{Def F_l}, we do the inverse Fourier transformation to get the probability density, and integrate it over $x$ to get the probability distribution.
Defining 
\bea
\Phi (x)\equiv \int _{-\infty }^x\frac{\text{dq}}{\sqrt{2\pi }} e^{-\frac{1}{2}q^2}
\eea
to be the error function. 
We obtain:
\bea\label{Edgeworth series}
F_l(x) = \Phi(x) + \sum_{j=1}^\infty \frac{q_j(-\partial_x)}{l^{j/2}} \Phi(x) .\eea
In general $q_j$ is a complicated polynomial, which can be compute to all orders. Here we write explicitly the first few terms: 
\begin{equation}
\begin{aligned}
%q_1(-\partial_x) &=-\frac{\kappa_3(\partial_x)^3}{6\kappa_2^{3/2}} \\
q_1  &=-\frac{\kappa_3(\partial_x)^3}{6\kappa_2^{3/2}} \\
q_2 &=\frac{\kappa_4(\partial_x)^4}{24\kappa_2^2} + \frac{\kappa_3^2(\partial_x)^6}{72\kappa_2^3} \\
q_3 &=-\frac{\kappa_3^3(\partial_x)^9}{1296\kappa_2^{9/2}}-\frac{\kappa_3\kappa_4(\partial_x)^7}{144\kappa_2^{7/2}}-\frac{\kappa_5(\partial_x)^5}{120\kappa_2^{5/2}} \\
&-\frac{(\partial_x)^3}{6}(\xi_2-\frac{3\xi_2}{2\kappa_2})
\end{aligned}
\end{equation}
Comparing the above result with the usual Edgeworth series \cite{EdgeworthKolassa,Edgeworth}, we find the first two terms are the same as those appearing in the Edgeworth expansion for $l$ independent measures with cumulants $\kappa_i$, the correction from the boundary term $\chi_0$ only effects the third and higher order terms. 

It is important to note that the parameters $\kappa_i,\xi_i$ are, in principle, measurable. For example, $\kappa_2,\xi_2$ can be obtained from the total noise in the measurement of $l$ and $l+1$ spins as  $\kappa_2=\langle  \Delta {\cal S}_{l+1}^2\rangle-\langle \Delta {\cal S}_{l}^2\rangle$ and $\xi_2=(l+1)\langle \Delta {\cal S}_{l}^2\rangle-l \langle \Delta {\cal S}_{l+1}^2\rangle $, where $ \Delta {\cal S}_l\equiv {\cal S}_l-\langle {\cal S}_l\rangle$. 

%It is important to note, that we do not assume $\chi_1(\lambda;l)$ to be a valid probability distribution, we only need that $log(\chi_1(\lambda))$ has a Taylor expansion, otherwise the Edgeworth series breaks down.  \\

Alternatively, by considering $\chi_0(\lambda)$ as a differential operator acting on the Fourier transform of $\chi_1$, and combining the Taylor series for $\chi_0(\lambda)=\sum_{k=0}^{\infty} {f_k\over k!} \lambda^k$ with the Edgeworth series for $\chi_1^l$, we may write explicitly
\bea\label{prob explicit sum}
\tilde{F}_{l}\!=\!\text{Prob}\!\left(\!\frac{{\cal S}_l-l \kappa _1}{\sqrt{\kappa_2 l}}\leq x\!\right)\!\!=\!\sum_{L=0}^{\infty }\!\frac{1}{l^{L/2}}\!\!\sum_{m=0}^L\!\!\frac{i^m\!f_m {\cal G}_{L-m,m}}{{
\kappa _2}^{m/2} m!}
\eea
%\bea
%\(\phi _n(x)=\sum_{k=0}^{\infty }\frac{{\cal Q}_k(x)}{n^{k/2}}
%\eea
where ${\cal G}_{k,m}(x)$ is given by ${\cal G}_{0,m}=(-\partial_{x})^{m}\Phi(x)$:
$$ 
{\cal G}_{k,m}=\!\!\!\!\!\underset{{\tiny\begin{array}{c}\{p_1,\text{..}p_k\}\in \mathbb{Z}_+^k \\ \Sigma  l p_l=k ; \Sigma  p_l=j \end{array}}}{\Sigma} \!\!\! \frac{(-\partial_x)^{k+m+2j}\Phi(x)}{p_1!..p_k!}\!\left(\!\frac{\kappa_3}{3!}\!\right)^{p_1}\!\!..\!\left(\!\frac{\kappa_{k+2}}{(k+2)!}\!\right)^{p_k}\!\!.
$$
We can now compute $F_l$ by using the expression \eqref{prob explicit sum}, combined with:
\begin{eqnarray*} &
F_l(x)=
\text{Prob}\!\left(\frac{{\cal S}_l-l \mu(l)}{(var({\cal S}_l)/l)\sqrt{l}}\leq x \right)=\nn & \text{Prob}\!\left(\!\frac{{\cal S}_l-l \kappa _1}{\sqrt{\kappa_2 l}}\!\leq \!{ \frac{1}{\sqrt{1+{\xi_2\over  l \kappa_2 }}}} x-\!\frac{\xi_1}{\sqrt{l \kappa _2}}\!\right)\!=\!\tilde{F}_l\left(\!{ \frac{1}{\sqrt{1+{\xi_2\over  l \kappa_2 }}}} x\!-\!\frac{\xi_1}{\sqrt{l \kappa _2}}\!\right)
\end{eqnarray*}

To illustrate these ideas, let us consider the following spin 1 MPS, given by the properly normalized matrices:
%\new  Can we write a concrete Hamiltonian for which this is a ground state?   \old
\[ A^+ \!\!= 
 \sqrt{\frac{1}{3}} \!\begin{pmatrix}
1 & 1 \\
0 & 0
\end{pmatrix} ;  A^0\!\!= \sqrt{\frac{1}{6}}\!
\begin{pmatrix}
-1 & 1 \\
1 & 1
\end{pmatrix} ; A^-\!\!= \sqrt{\frac{1}{3}}\!
\begin{pmatrix}
0 & 0 \\
-1 & -1
\end{pmatrix}
\] \\
Plugging these matrices in the definition \eqref{def of E(lambda)} we find that:   \\
\[ E(\lambda) = \frac{1}{6}  
\begin{pmatrix}
2e^{i\lambda}+1 & 2e^{i\lambda}-1 & 2e^{i\lambda}-1 & 2e^{i\lambda}+1 \\
-1 & -1 & 1 & 1 \\
-1 & 1 & -1 & 1 \\
2e^{-i\lambda}+1 & 2e^{-i\lambda}+1 & 2e^{-i\lambda}+1 & 2e^{-i\lambda}+1
\end{pmatrix} \]
At $\lambda=0$ we find that the largest eigenvalue is 1,  and that it is separated by a gap from the next eigenvalue $1/3$ . 

\begin{figure}
 \centering
\includegraphics[width = 0.45\textwidth]{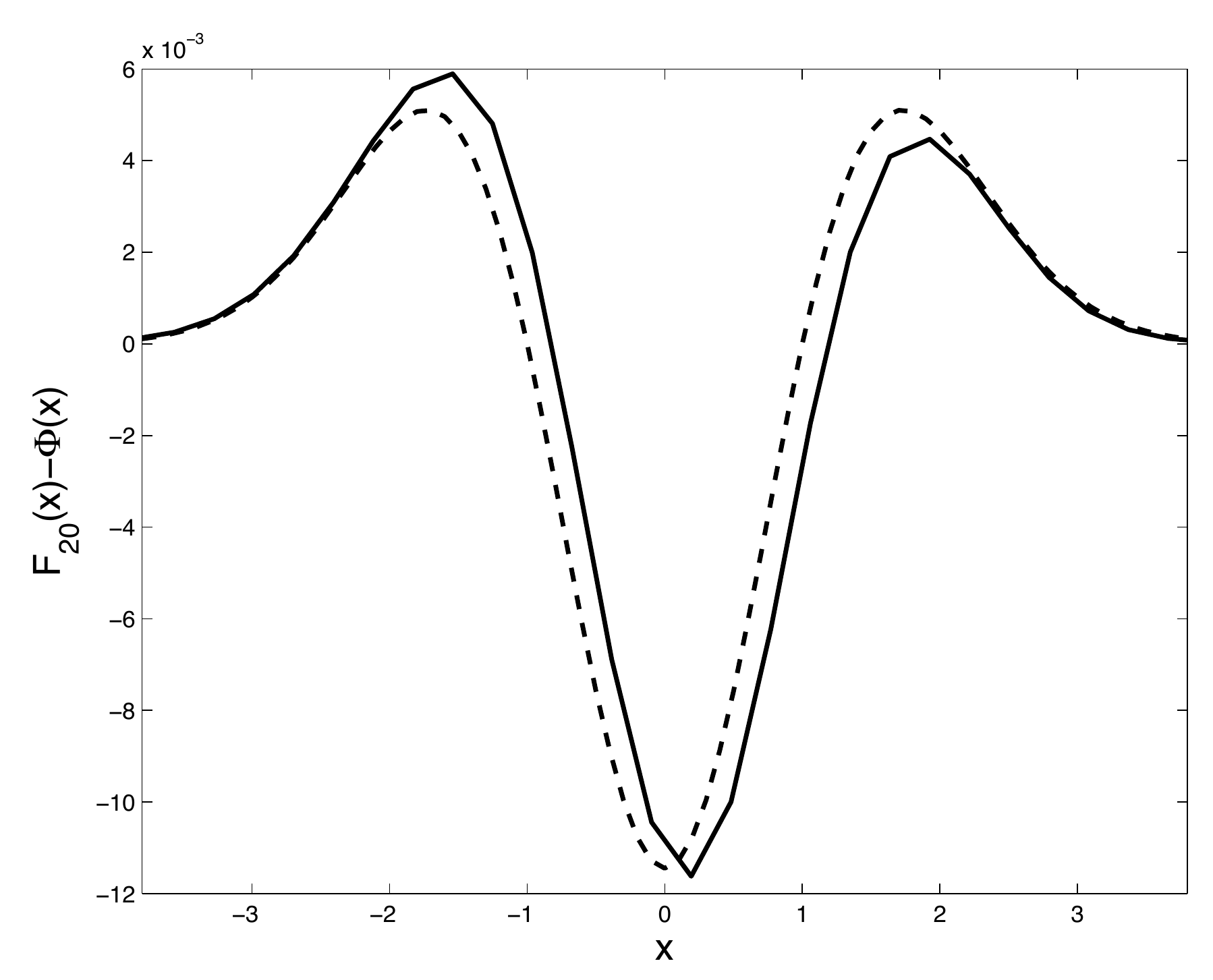}
\caption{Correction to the central limit distribution (i.e,$F_l(M)-\Phi(M)$). Solid line represent the exact result by calculating the probability distribution, dashed line shows the first order correction using Edgeworth series.}\label{Comparison}
\end{figure}

\begin{figure}
\includegraphics[width=0.45\textwidth]{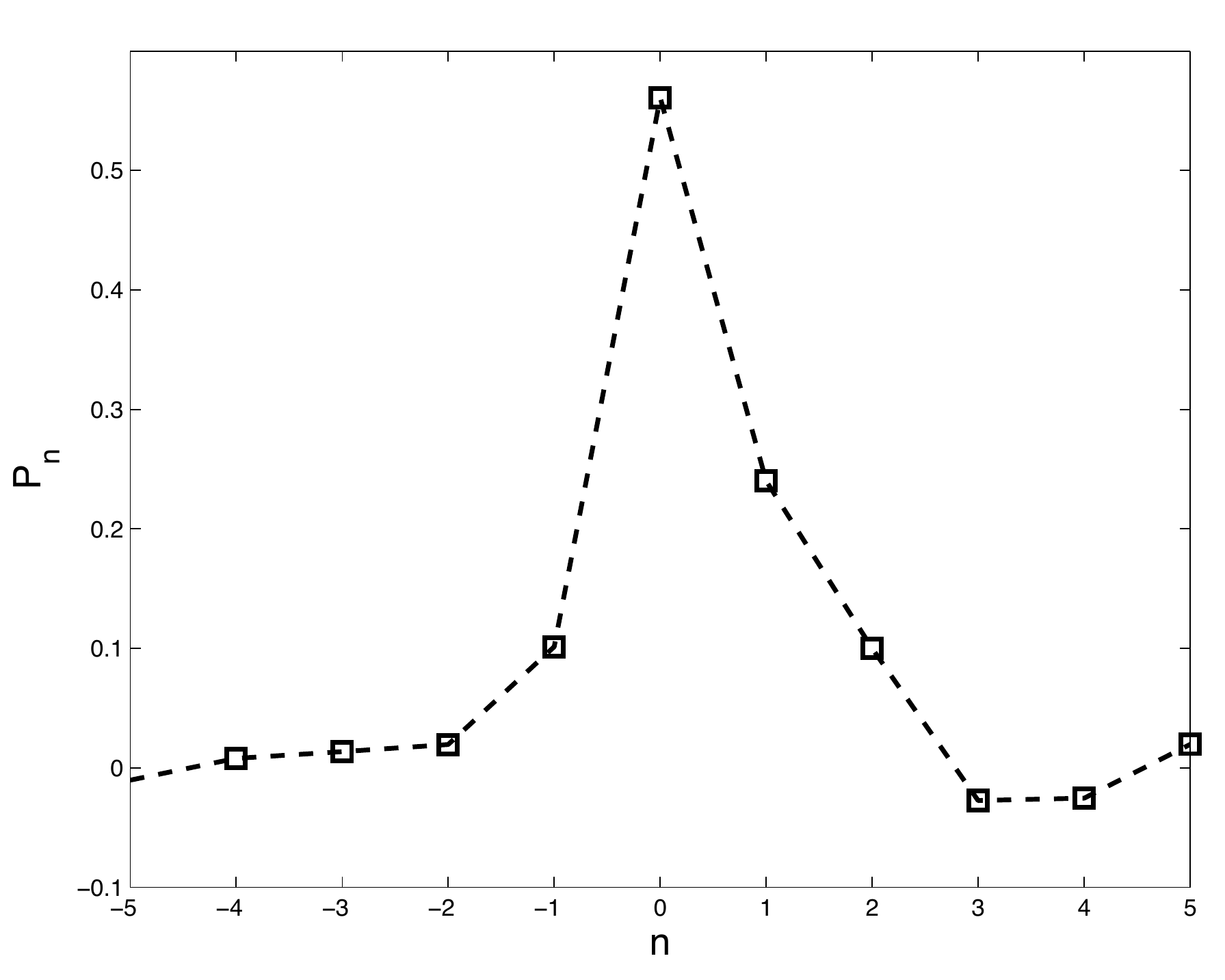}
\caption{First few Fourier components of $\chi_1$, showing small but finite negative pseudo-probabilities.}\label{fig negative probs}
\end{figure}

In Fig \ref{Comparison}. we compare the probability distribution for magnetization computed numerically from the ground state wave function, with the probability distribution obtained from our Edgeworth series \eqref{Edgeworth series}.
%Assume we have an infinitely long chain and we measure a block of $l=20$ spins. 
Here, the exact probability distribution $F_l(M)$ was found numerically by doing an inverse Fourier transformation of eq. \eqref{counting statistics chi}. 
In the example depicted in Fig \ref{Comparison}, for a block of 20 spins, it is evident that the Edgworth series works extremely well, capturing the essence of the correction already at first order. We also exhibit the pseudo-probabilities in Fig. \ref{fig negative probs}, computed according to eq. \eqref{pseudo probs}.

%We can also do this using the Edgeworth expansion, and fig(1) shows the correction of $F_l(M)$ from the error function $\Phi$.
%\subsection{Example: The AKLT state}
Next, we consider an example where the full counting statistics is not described by a Gaussian of finite width, although the system is gapped. In this example, the variance {\it per site} actually vanishes as $1/l$. 
Consider the so called  Affleck, Lieb, Kennedy and Tasaki (AKLT) state. The AKLT state \cite{AKLT1988} is a unique ground state of the AKLT Hamilitonian. It has a "string", instead of local, order parameter, and also fractionalized edge excitations. See \cite{AKLTtopology, Pollmann10} for a more recent study. The AKLT Hamiltonian is,
\[ \hat{H} = \sum_j \{\vec{S_j}\vec{ S_{j+1}} + \frac{1}{3}(\vec{S_j}+\vec{S_{j+1}})^2 \} . \]
The ground state has an MPS form, according to \cite{Schollwock11}
\[ A^+\!=\! \sqrt{\frac{2}{3}}
\begin{pmatrix}
0 & 1 \\
0 & 0
\end{pmatrix} ; A^0\!=\!\sqrt{\frac{1}{3}}
\begin{pmatrix}
-1 & 0 \\
0 & 1
\end{pmatrix}  ; A^-\!= \sqrt{\frac{2}{3}}
\begin{pmatrix}
0 & 0 \\
-1 & 0
\end{pmatrix}
\]
which gives:
\[ E(\lambda) = \frac{1}{3}
\begin{pmatrix}
1 & 0 & 0 & 2e^{i\lambda} \\
0 & -1 & 0 & 0 \\
0 & 0 & -1 & 0 \\
2e^{i\lambda} & 0 & 0 & 1
\end{pmatrix}. \]
Computing the eigenvalues of $E(\lambda)$, we see that they are $1, -\frac{1}{3}, -\frac{1}{3}, -\frac{1}{3}$, independent on $\lambda$. In the limit of $N, l \rightarrow \infty$, $e^S(\lambda, l, N) \rightarrow \frac{1+cos\lambda}{2}$, we find that the counting statistics does not depend on $l$. 

This result reflects the topological nature of the AKLT state, the total spin of the block depends only on the "edge modes" which are the only ones which are allowed to fluctuate.  

In general, the absence of scaling of fluctuations, appears whenever we have a correspondence of the form
\bea
E(\lambda)=V(\lambda)E(0)V^{-1}(\lambda)
\eea
For some matrix $V$. In such a case, $\chi_1(\lambda)\equiv 1$, and the entire contribution comes from the edge $\chi (\lambda ;l)=\chi_0(\lambda)$.  %\langle L(0)|V(\lambda)|R(0)\rangle \langle L(0)|V^{-1}(\lambda)|R(0)\rangle,  \eea
In this case  $\log \chi(\lambda)$ is clearly not extensive in the block size $l$. %The Edgeworth expansion since it is not suitable to describe a delta function like probability distribution.

To summarize, Matrix product states supply a very natural class of probability distribution which are not IID, but are still quite amenable to treatment and computation. In this paper we have studied the full counting statistics of spin in matrix product states. We explored the finite size correction to the Gaussian counting statistics expected on large scales. We showed how an Edgeworth type series may be used describe these asymptotic corrections and checked it numerically on an explicit example. Finally, we showed that in special cases, such as the AKLT model, the fluctuations in the system  do not scale linearly with system size, and the Edgeworth description is not valid whenever it is based on variance and mean which were measured on any finite size block. 

{\bf Acknowledgments:} Financial support from NSF CAREER award No. DMR-0956053 is gratefully acknowledged. 
\bibliographystyle{apsrev}
\bibliography{MPScounting.bib}

\end{document}